\documentclass[11pt]{article}
\begin{document}
\newcommand{\ben}{\begin{eqnarray}}
\newcommand{\een}{\end{eqnarray}}
\newcommand{\la}{\label}
\title{Schr\"odinger Quantization Problem and Neutron Stars}
\author{Plamen P.~Fiziev \thanks{E-mail:\,\, fiziev@phys.uni-sofia.bg}}
\maketitle
\centerline{Department of Theoretical Physics,
Sofia University,}
\centerline{ 5 James Bourchier Boulevard, BG-1164, Sofia, Bulgaria}
\centerline{and}
\centerline{The Abdus Salam International  Centre for Theoretical Physics,}
\centerline{Strada Costiera 11, 34014 Trieste, Italy}
\begin{abstract}
We are discussing the possibility to find a proper unique
conditions for an experimental study of the Schr\"odinger quantization
problem  in the neutron stars physics. A simple toy model
for physically different quantizations is formulated and  a possible
physical consequences are derived.

\end{abstract}

\sloppy
\scrollmode

\section{Introduction}

The quantization of the classical mechanical systems is still an
open  problem. It was raised as a {\em physical} problem already in the
pioneering article on quantum mechanics by Ervin Schr\"odinger
\cite{Sch} in the following form. Suppose we are given a classical
particle with mass $m$ in a potential field $V({\bf r})$. Its
classical Hamiltonian is
\ben H={{\bf p}^2\over{2m}}+V({\bf r}).
\la{Hcl} \een
The problem is how {\em to find} a quantum operator $\hat
H$ which corresponds to the classical Hamiltonian (\ref{Hcl}) and which is
needed to write down the Schr\"odinger equation: \ben i\hbar\,
\partial_t \psi({\bf r},t)= \hat H\, \psi({\bf r},t), \la{SchE}
\een

In the coordinate representation we know the position operator
$\hat{\bf r}={\bf r}$ and the momentum operator $\hat{\bf
p}={\hbar\over i}\nabla$. Unfortunately this is not enough to find
the quantum Hamiltonian $\hat H$. For this purpose we have to
solve, in addition, the operator ordering problem for the
non-commutative canonically conjugated variables $[\widehat p_i,
\widehat x_j]={\hbar \over i}\delta_{ij}$. The superposition
principle tells us that $\hat H=\widehat{{\bf
p}^2\over{2m}}+\widehat{V({\bf r})}$, but what are $\widehat{{\bf
p}^2}$ and $\widehat{V({\bf r})}$ is not known.

Schro\"dinger has stressed that, for example, we can write down
the classical quantity $p_i^2$ in a classically equivalent form
${1\over{f({\bf r})}}p_i f({\bf r})^2 p_i{1\over {f({\bf
r})}}\equiv p_i^2$ with an arbitrary function $f({\bf r})\neq
0,\infty$. Then the canonical "hat-quantization" will give a
corresponding quantum operator $\widehat{{1\over {f({\bf
r})}}}\,\widehat{p_i}\, \widehat{f({\bf r})^2}\,\widehat{ p_i}\,
\widehat{{1\over {f({\bf r})}}}= {\widehat p_i}^2+
{\hbar^2}\partial_i^2 f({\bf r})/f({\bf r}) = {\widehat p_i}^2 +
{\cal O}(\hbar^2)$. As a result, in coordinate representation we
will have a quantum Hamiltonian

\ben \widehat H=-{\hbar^2\over{2m}}\Delta +{\hbar^2\over{2m}}{{\Delta
f({\bf r})}\over f({\bf r})} + V({\bf r}). \la{Hq} \een

The real quantization problem is not limited to this simple example.
A general form of this problem is related with the fact that we do not
know how to ``quantize'' the simple classical number $1$, when considered
as a dynamical ``variable''.
This becomes transparent, if we write down the number $1$ in
a form of a product $1\equiv f_1(x)g_1(p)f_2(x)g_2(p)...f_n(x)g_m(p)$
of an arbitrary well defined functions $f_1(x),...,f_n(x)$ and
$g_1(p),...,g_m(p)$, such that
$f_1(x)f_2(x)...f_n(x)\equiv 1$ and  $g_1(p)g_2(p)...g_m(p)\equiv 1$.
These functions have to be introduced {\em at classical level.}
Therefore they certainly are independent of the Planck constant $\hbar$
and can depend only on the classical macroscopic
parameters of the system at hand. Now, ``putting the quantum
hats'' on the letters, we obtain a nontrivial quantum operator,
which corresponds to the simple number
$1$:
 $$\widehat I:= f_1(\hat x)g_1(\hat p)f_2(\hat x)g_2(\hat p)...
f_n(\hat x)g_m(\hat p).$$
This operator has a limit $\widehat I\to 1,\,\,\,\hbox{for}\,\,\,\hbar\to 0$

It is clear, that  we have infinitely many such quantum operators
$\widehat I$ and we can put any of them at any position in the
expression for every quantum quantity, obtained from some
classical one by the use of the correspondence principle. For this
purpose one has to put in the corresponding classical quantity the
number $1$, as a multiplier, at different  positions and then,
during the ``quantization'', we have to replace these numbers by
some of their quantum representatives $\widehat I$. Hence,
actually there exist no physical problems for which one can ignore
the above Schr\"odinger quantization problem. We need to use some
additional principles, which are independent of the correspondence
principle, to be able to fix the right quantization in a given
physical problem. After all, the term ``right quantization'' can
have only one meaning: the quantization, which reflects in an
adequate way the real properties of the physical objects.

\section{Some of the Mathematical Problems,
Related to the Schrodinger Quantization Problem}

It is not easy to formulate in a mathematically correct way the
Schr\"odinger quantization problem, when we consider this problem
in the above very wide framework. Here we shell only mark some of
the existing problems and approaches.

The first mathematical problem which one must consider, is the
fixing of the Hilbert space of the quantum states. This is known
to be not a trivial problem and many of the formal operators,
obtained by the correspondence principle may turn to be not well
defined {\em self-adjoint} operators on this Hilbert space. For
example, in the simple infinite wall problem the quantum momentum
$\widehat p$ is not well defined self-adjoint operator. A direct
way to see this is to take into account that the quantum operator
$\exp({i\over\hbar}\,a\,\widehat p)$ represents a translation in
the coordinate space. This translation will drop out of the any
compact finite interval $x\in [x^\prime,x^{\prime\prime}]$ some
part of the quantum states, which in this problem are defined as a
functions on this fixed interval: $\psi(x)\neq 0$ at least for
some values of $x\in [x^\prime,x^{\prime\prime}]$, and
$\psi(x)\equiv 0$ for $x\bar\in [x^\prime,x^{\prime\prime}]$.

In contrast, the requirement to have Hermitian operators with
respect to some fixed measure in the corresponding quantum space
of states, imposes a very weak restriction on the form of the
quantum operators and may be easily fulfilled.

Quite more restrictive result is the Van Hove theorem \cite{VanHove}.
It states that it is impossible to find a quantization procedure, which
transforms the Poisson algebra of the classical
quantities onto a quantum one, changing
the Poisson brackets with proper quantum commutators.
This result was the starting point of developing of the
geometrical quantization \cite{Sims}, where the important
notion of a polarization of the classical phase space was introduced.

At first glance it seems that the Feynman path integral approach
to quantum mechanics \cite{Feynman} is able to solve the
Schr\"odinger quantization problem, since in this approach only
classical notions are used. As it was shown in \cite{Beresin},
this is not the case. The very Feynman path integral depends on
the discretization procedure and this is complete equivalent to
the operator ordering problem in the above canonical quantization
procedure. A various discretization, which lead to $\hat p-\hat x$
quantization, $\hat x-\hat p$ quantization, Weyl quantization,
Wick quantization, etc., can be easily found. Hence, the ambiguity
in the path integral is of the same nature as the uncertainty in
the canonical Schr\"odinger quantization.

After all it turns out that the Feynman path integral {\em on the
classical phase space} can throw some new light on the
quantization problem. As shown in \cite{Fiziev}, the result of the
summation on the virtual paths on the classical phase space
strongly depends on the choice of the class of the paths, which we
include in the Feynman sum. This choice can be described {\em
completely} in terms of classical mechanics, using different
maximal sets of classical first integrals in involution. The
choice of such set of local classical first integrals fixes the
quantization. This observation may lead to a real progress in the
understanding of the quantization problem, since now the very
ambiguity of the quantization procedure is described completely
only in terms of the classical mechanics. This approach is deeply
related with Van Hove theorem and geometric quantization. Another
advantage is that the procedure for calculation of the Feynman
path integral, based on this approach, uses the Volterra's
multiplicative integral. This permits us to overcome the
restriction to consider {\em only} Gaussian measures in the path
integrals. Unfortunately, at present this approach is not
developed enough and suffers from some specific technical
difficulties.

There exist another widespread point of view on the operator
ordering problem. If one accepts a totally quantum philosophy, one
can think that the quantum operators are the primary given
objects. Then we {\em postulate} the form of  quantum Hamiltonian
$\widehat H$, which is considered {\em as known} without any need
to quantize classical quantities. Then the correspondence
principle may be used to establish a correspondence of the {\em
given} quantum operators to the classical quantities.

In this approach a pure technical problem appears: how to write
down the given quantum Hamiltonian $\widehat H$ in the terms of
the given quantum operators of the positions -- $\widehat x_i$ and
of the momenta -- $\widehat p_i$. This formal problem admits a
variety of  mathematical solutions. For example, one can write
down the quantum Hamiltonian in the following forms: $\widehat
H=\Sigma_{m,n}H^{px}_{mn}\,\widehat p^n\widehat x^m \equiv
\Sigma_{m,n}H^{xp}_{mn}\,\widehat x^m\widehat p^n \equiv
\Sigma_{m,n}H^{\{px\}}_{mn}\,\{\widehat p^n\widehat x^m\} \equiv
...$. Here the symbol $\{...\}$ stands for some kind of
symmetrization, for example, $\{\widehat p^n\widehat
x^m\}=\widehat p^n\widehat x^m+\widehat x^m\widehat p^n$. The
numerical coefficients $H^{px}_{mn}, H^{xp}_{mn},
H^{\{px\}}_{mn}...$ are different and depend on the Planck
constant. In this case the corresponding principle says that
replacing the operators $\widehat x^m$ and $\widehat p^n$ with
their classical counterparts $ x^m$ and $p^n$ we will obtain the
classical Hamiltonian $H(p,x)$ in the limit $\hbar \to 0$, no
mater which of the above representations of the quantum
Hamiltonian we use.

It is clear that this point of view has noting to do with the Schrodinger
quantization problem. It is designed to solve only the above formal mathematical
problem.

In the present article we will not pay attention to the last formal
mathematical problem. Instead we shall make an attempt to consider
the Schrodinger quantization problem as a real physical problem.

\section{The Physical Dificulties in the Schrodinger Quantization Problem}

The real { physical} difficulties originates from the fact, that
the differences between different quantizations of a given
classical quantity may be of high order with respect to the Planck
constant $\hbar$. This means that the deviations from a chosen
quantization of this quantity can be extremely small and behind
the real experimental abilities for  measurements. Therefore at
present we have no good real experimental evidences in favor of
some definite quantization rule, or in favor of some more general
{\em physical} theory, which uses simultaneously different
quantization rules for different physical purposes.

\subsection{The Harmonic Oscillator}

We shall illustrate the general situation using a simple toy
model. Consider the one-dimensional harmonic oscillator with
classical Hamiltonian \ben H = {1\over {2m}}p^2 +{m\over
2}\omega^2 x^2. \la{Hoscc}\een Let us choose in the formula
(\ref{Hq}) a cutting-like function $f(x)=e^{-x^2/2a^2}\to 1$ for
$a\to \infty$, with some macroscopical parameter of length $a$.
Then the corresponding quantum operator may be written in the
form: \ben
 \widehat H = {{-\hbar^2}\over {2m}}\partial_x^2 +
{m\over 2}\bar\omega^2 x^2 -{\hbar^2\over {2m a^2}} \la{Hoscq}\een
with some corrected circular frequency \ben \bar\omega = \omega\,
\sqrt{1+ {{\hbar^2}\over{\omega^2m^2a^4}}}. \la{omegah} \een

Now it becomes clear that the use of the function $f(x)=e^{-x^2/2a^2}$
in the above nonstandard quantization procedure has a simple physical meaning.
This function introduces some specific {\em quantum constraint}
on the motion of the quantum particle via some specific quadratic
quantum wall: $V_{quant}={{\hbar^2}\over{2ma^2}}\left(x^2/a^2-1\right)$.
As a result, outside some domain, defined by the parameter $a$, i.e.,
for a big values of the coordinate $x>>a$,
the wave function now decreases exponentially,
(almost)independently of the behavior of the real physical
potential $V(x)$. This potential may influence only the details of
the exponential decay of the wave function, if $V(x)$  is limited from
the below for $|x|\to \infty$,
i.e. if $V(x)\geq const > -\infty$ for $|x|\to \infty$.
When the parameter $a$ increases, the domain,
where the wave function may oscillate, increases, too.
In the limit $a\to \infty$ this specific quantum constraint disappears.

The spectrum of the Hamilton operator (\ref{Hoscq}) is described by the formula
\ben E_n(a)=\left(n+{1\over 2}\right) \hbar\bar \omega -{\hbar^2\over {2m a^2}}=
\nonumber \\
E_n(\infty) -{\hbar^2\over {2m a^2}} +
\left(n+{1\over 2}\right) {\hbar^3\over {2\omega m^2 a^2}}+{\cal O}(\hbar^4).
\la{En}\een

As we see,  the standard  levels $E_n(\infty)=\left(n+{1\over 2}\right)\hbar \omega$
are shifted by the amount of $\Delta E_n(a)= -{\hbar^2\over {2m a^2}}$ and,
in addition, we have an increase of the relative distance between the  levels,
described by the dimensionless quantity
\ben {{\delta E_n(a)}\over{E_n(\infty)}}={1\over 2}\,\delta^2=
{1\over 2}\,{\hbar^2\over {\omega^2 m^2 a^4}}
\la{deltaEn}\een
in the lowest order with respect to $\delta$. The dimensionless parameter
$\delta={\hbar\over {\omega m a^2}}$ controls the deviation of
our nonstandard quantization of the harmonic oscillator
from the ``standard'' one.

The generalization of this example for the D-dimensional isotropic oscillator is obvious.
Choosing the function $f=\exp\left(-{1\over{2a^2}}\Sigma_{i=1}^D\,(x_i)^2\right)$,
we immediately obtain
the quantum Hamiltonian:
\ben
 \widehat H = {{-\hbar^2}\over {2m}}\Delta_D +
{m\over 2}\bar\omega^2 x^2 -D{\hbar^2\over {2m a^2}}
\la{HDoscq}\een
with the same circular frequency $\bar\omega$, defined by the formula (\ref{omegah}).

\subsection{The Hydrogen Atom}

It is well known, that  there exist a simple correspondence
between the spectrum of the D=2 isotropic oscillator and the
spectrum of the hydrogen atom, see for example \cite{Schwinger}.
One easily obtains that when one  uses in this correspondence the
quantum Hamiltonian (\ref{HDoscq}), instead of the usual one (with
$a=\infty$), one will arrive at  the following corrected formula
for the spectrum of the hydrogen: \ben
E_n:=-{{R_\infty}\over{n^2\left(1+\delta^2\right)}}
\la{Ehydrogen}\een where $R_\infty=13.605\,691\,72(53)\, eV$
(Uncert. (ppb): 39) is the standard Rydberg constant. Hence, such
unusual quantization of the hydrogen atom is equivalent to a
correction of the Rydberg constant with a relative decrease
$\delta  R_\infty / R_\infty=-\delta^2$.

Now, making use of the standard formulas for the hydrogen, we
obtain: \ben \delta= \left({{a_B}\over a}\right)^2 \la{deltaB}
\een where $a_B\approx 0.529177  \times 10^{-8}\,\hbox{cm}$ is the
Bohr radius. Hence, for a macroscopic length $a\sim 1\,\hbox{cm}$
we will have $\delta\sim 10^{-16}$. This will lead to relative
correction of the Rydberg constant of order of magnitude
$\delta^2\sim 10^{-32}$ -- too far from any experimental abilities
in any foreseen future.

The general rule for inclusion of the correction for our nonstandard quantization
in different quantum qantities, related with hydrogen, is very simple:
one has to make only the replacement of the principal quantum number
$n \to n\,\sqrt{1+\delta^2}$. Then we can immediately see the influence of this nonstandard
quantization on the Lamb shift. Using standard notations \cite{Schwinger} we obtain:
\ben
\delta E_{n,0}={4\over{3\pi}}{{\alpha^3 Z^4}\over{n^3}}
\log\left({{m_{el}c^2}\over{\Delta E}}\right)
{{e^2}\over{a_0}}\left(1+\delta^2\right)^{-3/2}.
\la{Lamb}
\een

Hence, the relative deviation in the Lamb shift due to our nonstandard quantization
procedure will be of order of magnitude $-{3\over 2}\delta^2 \sim -10^{-32}$.
Thus we see that although the measurements of the Lamb shift are at present between
the experimental checks of quantum mechanics with a greatest possible precision,
these measurements are not able to see such small deviations from the
``standard'' canonical  quantization of the hydrogen.

\section{Schr\"odinger Quantization Problem and Neutron Stars}

As we sow in the previous section, from experimental point of view
the Schr\"odinger quantization problem seems to be hopeless issue.
The corrections to the basic quantum observables are extremely
small for systems with small number of degrees of freedom. At the
same time the formula (\ref{HDoscq}) shows explicitly that the
common shift of the levels  is proportional to the number of the
degrees of freedom. Obviously the same conclusion is valid in the
general case, too, because we are quantizing independently all
degrees of freedom and these enter the classical Hamiltonian
additively. For systems with a huge amount of degrees of freedom
the huge number D can compensate the extremely small quantization
correction $\sim \delta ^2$ for a single degree of freedom. Then
the physical effects may be significant. Hence, to be able to see
the effects of the possible different quantization procedures we
need some physical system with a huge number of degrees of freedom
$D\sim \delta^{-2}$ and in addition this system has to be in a
{\em coherent} quantum state.

The ideal known physical systems of this sort are the neutron
stars. At present their behavior is observed with a great details.
See for an additional information, for example, the review
articles \cite{Becker, Comer} and the references therein. Indeed,
the neutron stars have some $10^{57}$ neutrons in them in a
coherent superfluid quantum state. The typical density of matter
in the neutron stars is $\,\,\approx 2.8 \times 10^{14}$
${g\,cm^{-3}}$. In addition, in the neutron stars we have a
magnetic fields of order of magnitude $\sim
10^{12}\,\,\hbox{Gauss}$ and a rotational energy $\sim 2\times
10^{49}\,\,\hbox{erg}$. These extremal physical conditions make
the neutron stars a unique astrophysical laboratories.

The typical radius of the neutron stars is of order of $3\times
10^{5}\,\,\hbox{cm}$. Hence, if we accept the macroscopic
parameter $a$ to be of order of magnitude of the star's radius, we
will have for the neutron star $\delta\sim 10^{-26}$. If the
quantum corrections due to the change of the quantization
procedure (like that for the levels of the hydrogen, or for the
Lamb shift) are of order of $\delta^2\sim 10^{-52}$, this
extremely small number can be compensated by the huge number of
the degrees of freedom of the neutrons in the star. Then for the
parameter $D\,\delta^2$, which controls the possibility for real
observation of the effects of different quantization procedures,
we obtain $D\,\delta^2\sim 10^{5}$ -- an unique big number, in
comparison with the corresponding extremely small values of the
corrections of this type in all available Earth-laboratory
experiments.

\section{Concluding Remarks}

Ofcourse the above consideration is a pure qualitative speculation.
Although it is not based on some deep theoretical analysis, it shows that,
in principle, an attempts to look for some phenomena in neutron-star physics,
which are related with the Schr\"odinger quantization problem,
may have a good physical ground. If we are extremely lucky, it could happen
that the macroscopic parameters like the parameter $a$ in the simple toy model,
considered in the present article, are much smaller then the radius of the star.
This may increase essentially the important factor $D\,\delta^2$.

On the other hand, many of the observed phenomena in
neutron-star's physics are not well understood at present. We
shall mansion only two of them: the mechanism of star's radio
emission \cite{Becker} and the instabilities of the oscillations
of the stars -- the so called glitch phenomenon in pulsars -- a
sudden spin-up of the star's crust \cite{Comer}. Ofcourse, a more
conventional models of these phenomena, without consideration of
quantization problem,  are at present in an intensive study.

It seems to us that neutron stars, which appeared at first as a pure theoretical
invention in the early 1930's on the ground of the quantum mechanics and gravity
in the independent articles by Landau and by Chandrasekhar, may help us once more
to reach a deeper understanding of the quantum mechanics and, especially, for a
discovery of a new physics behind the Schr\"odinger quantization problem.

It is not excluded, too, to look for some new effects, related to
the Schr\"odinger quantization problem in a more usual laboratory
conditions, studying the collective quantum phenomena like
superfluidity and superconductivity, and using the possible
interplay between microscopical quantities, like the Bohr radius
$a_B$, and the macroscopical ones, like the parameter $a$ in the
above consideration.

\vskip .5truecm

{\bf Acknowledgments}
\vskip .3truecm

I am deeply indebted to Professor S.~Randjbar-Daemi for his invitation to visit ICTP,
Trieste, where this article was completed.

My special thanks to Professor Ennio Gozzi for his kind help, support
and many stimulating discussions and comments on the quantization problem
during a long period of time.

I am thankful, too, to Professors G.~L.~Comer and N.~Anderson for useful
discussions on neutron star's physics in the context of the quantization problem
during the Conference on Sources of
Gravitational Waves, September 2003, Trieste.

This research was supported in part  by Scientific Found of
Sofia University, Grant Number 3305/2003,
by the University of Trieste, by INFN, and by the Abdus Salam International
Centre for Theoretical Physics, Trieste, Italy.

\end{document}